# We are IntechOpen,
# the world's leading publisher of Open Access books
# Built by scientists, for scientists

**4,700**
Open access books available

**121,000**
International authors and editors

**135M**
Downloads

**154**
Countries delivered to

Our authors are among the

**TOP 1%**
most cited scientists

**12.2%**
Contributors from top 500 universities

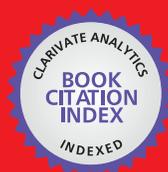

**WEB OF SCIENCE™**

Selection of our books indexed in the Book Citation Index
in Web of Science™ Core Collection (BKCI)

# Interested in publishing with us?
# Contact book.department@intechopen.com

Numbers displayed above are based on latest data collected.
For more information visit www.intechopen.com

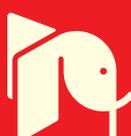

# Wind Power Variability and Singular Events

Sergio Martin-Martínez,
Antonio Vigueras-Rodríguez, Emilio Gómez-Lázaro,
Angel Molina-García, Eduard Muljadi and
Michael Milligan

Additional information is available at the end of the chapter

http://dx.doi.org/10.5772/52654

## 1. Introduction

There are several sources of wind power variability, including short-term (seconds to minutes) and long-term (hours of the day, or seasonal). The type of wind turbine, the control algorithm, and wind speed fluctuations all affect the short-term power fluctuation at each turbine. The blade of a Type 1 induction generator wind turbine experiences the tower shadow effect every time a blade passes the tower; the output of the Type 1 turbine commonly fluctuates because of this. Wind turbines Type 3 and Type 4 are equipped with power electronics and have a reasonable range of speed variation; thus, the impact of the tower shadow effect is masked by the power converter control. Because of wind speed fluctuations and wind turbulence, the power output also influences the output of the wind turbine generator.

However, in the big picture, the power fluctuation at a single turbine is not as important as the total power output of a wind power plant. The interface between the wind power plant and the power system grid is called the point of interconnection (POI). All the meters to calculate revenue, measure voltage and frequency, and measure other power quality attributes are installed at the POI. At the POI, all the output power from individual turbines is injected into the power grid. A wind power plant covers a very large area; thus, there are various diversities within each plant (e.g., wind speed, line impedance, and instantaneous terminal voltage at each turbine). The power measurement from a single wind turbine usually shows a large fluctuation of output power; however, because many turbines are connected in a wind power plant, the power fluctuation from one turbine may cancel that of another, which effectively rectifies the power fluctuation of the overall plant.

Additionally, many wind power plants are co-located in the same region where wind resources are excellent; thus, the spatial diversity among wind power plants contributes to a smoother output power of the region than the output power of an individual wind power plant.

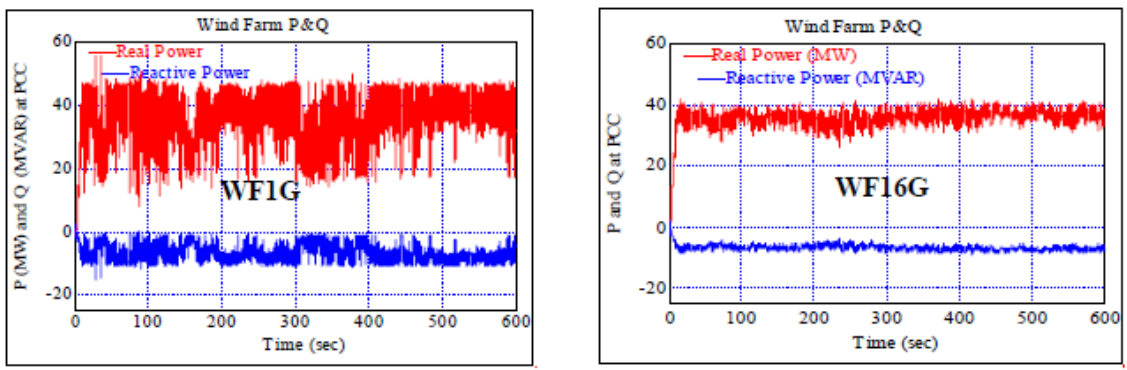

**Figure 1.** Real and reactive power output of a wind power plant; (a) Single turbine representation (b) Sixteen turbines representation

Figure 1(a) shows the fluctuation in the output power when the wind power plant is represented by a single turbine. Figure 1(b) illustrates the real and reactive power output when the wind power plant is represented by sixteen turbines. With greater wind diversity, as shown in Figure 1(b), the power fluctuation is smoother than that with less wind turbine representation, as shown in Figure 1(a) [1].

## 2. Wind speed variability

Overall wind variability consists of different fluctuating terms with different periods, depending on the sources. For instance, fluctuating term can be caused by day-night effects (e.g., the effects of sea breeze), because there are "quick" fluctuations in some minute periods.

Figure 2 shows a typical power curve of a variable speed wind turbine generator. In general, the operating wind speed is divided into different regions. In Region 1 and Region 2 (low to rated wind speed), a wind turbine is operated in variable speed at a constant pitch angle (typically 0°). The output power of the generator is low to rated power. The operation is optimum because the turbine is operated at the maximum performance coefficient Cp. When wind speed varies, the output power varies as the cube of wind speed variations. Once wind reaches its rated speed, the rotational speed also reaches the rated speed. This rotational speed must be limited by the pitch control to keep the rotor speed from a runaway condition and to limit the mechanical stresses of the wind turbine structure (tower, blades, and gearbox). The output power is limited to its rated value.

Thus, wind power output varies only when wind speeds are below rated value. Below rated wind speed, the rate of change of the output power $\left(\frac{dP}{dt}\right)$ is either positive or negative, de-pending on the direction of wind speed change. Above rated wind speeds, any fluctuations will be capped at rated by the pitch action, or $\frac{dP}{dt}=0$.

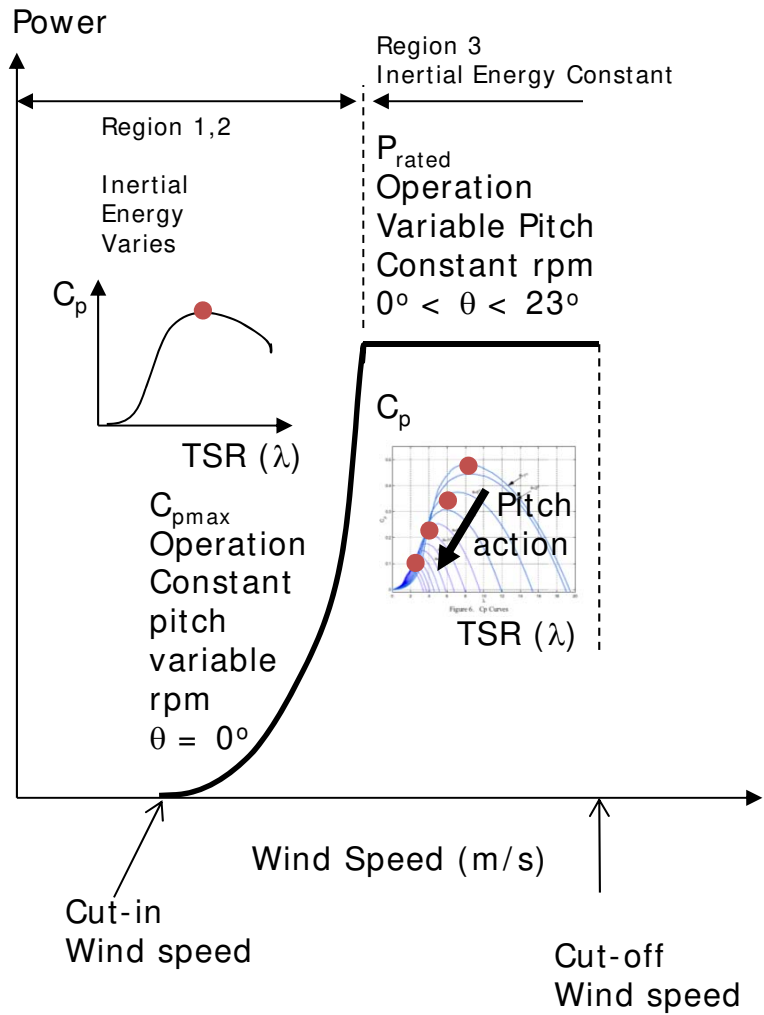

**Figure 2.** Power curve of a typical wind turbine generator

If the power controller is not properly designed, wind fluctuations may excite the mechanical resonance of the structure or gearbox, which may lead to mechanical failures of the wind turbine [2]–[3].

Spectral tools are often used to analyze wind speed variability because they make it possible to study different frequency fluctuation terms. The most popular one used for this purpose is the Power Spectral Density (PSD).

The PSD of a function is defined by the Fourier transform of its autocorrelation. PSD is therefore expressed in frequency domain. Its physical meaning is related to fluctuating kinetic energy on a certain frequency.

Van der Hoven [4] analyzed the PSD of horizontal wind speed, based mainly on measurements done at Brookhaven National Laboratory. Figure 3 shows such spectra, with two main peaks and a spectral gap between them. The first peak occurred at a period of around four days and was caused by migratory pressure systems of a synoptic weather map scale. The second peak occurred at a period of 1 minute because of a mechanical and/or convective type of turbulence. Van der Hoven's observations also showed some relation between the spectral gap shape and surface roughness under some circumstances. Additional analysis shows more complex spectra, especially over the ocean or in smooth terrains [5]–[9]; where there is an important contribution of mesoscale fluctuations combined with various phenomena such as convective longitudinal rolls [11] or cumulus clouds [12] that may contain considerable spectral density in the frequency range. At some other places, the gap was verified by experimental data [4], [13]–[14].

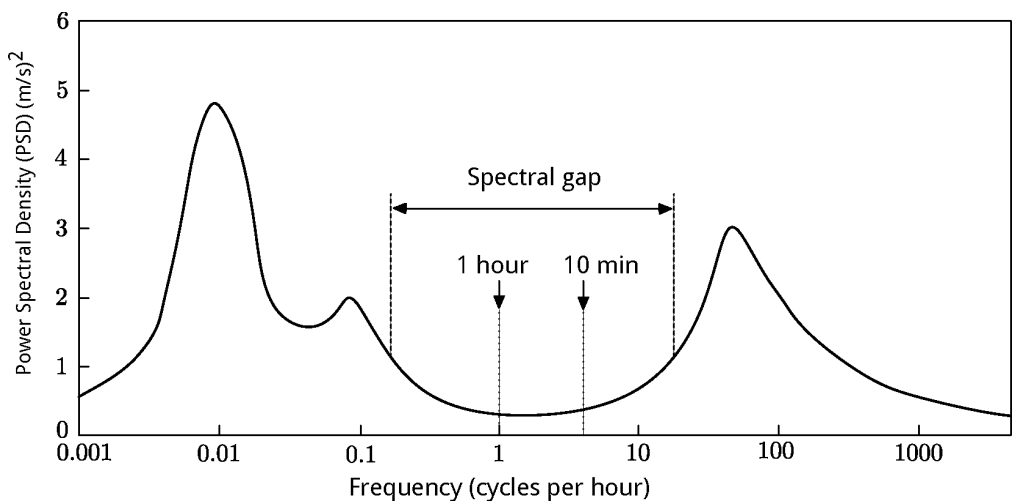

**Figure 3.** Spectrum estimated by Van der Hoven [4]

PSD spectrum is usually calculated by using short segments of data with similar atmospheric characteristics. The reduction or removal of the spectral gap introduces some difficulties on the analysis because microscale and macroscale are no longer separated. This limits the findings of particular atmospheric regimes lasting long enough to calculate a meaningful spectrum. Thus, some researchers are considering the use of more complex spectral tools based on time and frequency domain. For instance, the Hilbert-Huang transform has been used for analyzing wind fluctuations over the North Sea [8].

Wind speed variability is important with regard to power system management. An example of the significance of these power fluctuations is in Energinet.dk (the Danish Transmission System Operator). According to [15], Energinet.dk has observed that power fluctuations from the 160-MW offshore wind power plant at Horns Rev in West Denmark introduce several challenges to reliable operation of the local power system. The power fluctuations also contribute to deviations from the planned power exchange with the Central European Power System. Moreover, it was observed that the timescale of the power fluctuations was from tens of minutes to several hours.

Figure 4 shows the relation between time and geographical scales and the impacts affecting power system operation. Depending on the level of the wind power penetration, power fluctuations due to wind speed variability may influence the frequency regulation; transmission and distribution efficiency, and load flow; or even efficiency of the thermal and hydro power plants connected to the same grid.

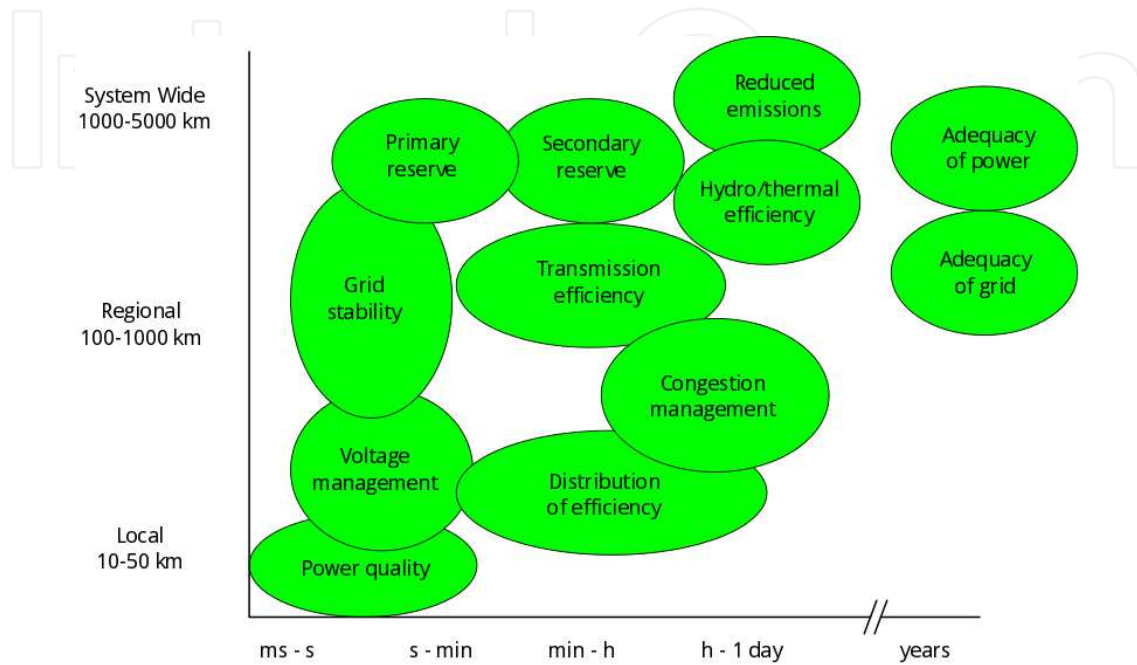

**Figure 4.** Time and geographical scales of power system issues [16]

## 3. From wind speed variability to regional wind power fluctuations

As highlighted above, from the point of view of power systems, it is important to analyze wind power fluctuations from tens of minutes to several hours. Therefore, it is important to study the relation between wind speed variability and wind power fluctuations.

Different operating regions give different rates of power fluctuations. The first stage in the conversion of wind power to electrical power is the smoothing and sampling effect produced by the size of the wind turbine rotor. In fact, some variations in wind speed at a single point within the rotor swept area are smoothed out when considering the entire blade length. Particularly, uncorrelated oscillations of wind speed are attenuated when considering several points within the rotor swept area. To analyze the correlation between such oscillations, spectral coherence is usually considered. Studies have been done of spectral coherence between the horizontal wind speeds within the rotor disk, showing a large smoothing of the (high-frequency) quickest variations [17]. To illustrate these effects, Figure 5 compares wind speed at a single point within the rotor swept area, with an equivalent wind speed over the rotor disk [18].

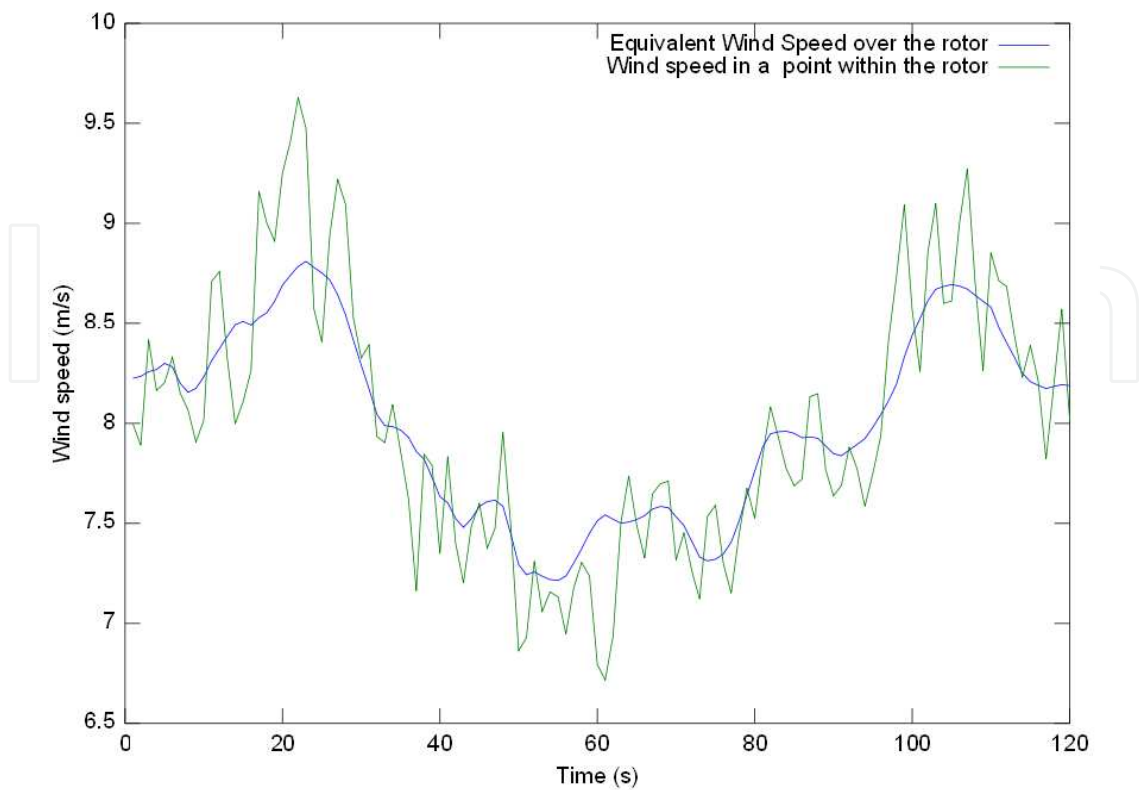

**Figure 5.** Comparison between wind speed in a single point of the rotor disk and the equivalent wind over the rotor disk

Moreover, the conversion of wind energy into electrical power is not linear. In a typical power curve (Fig. 2), there are wind speed ranges with different impacts on the conversion from wind speed variations to wind power fluctuations. Between cut-in wind speed and nominal wind speed, variability in power tends to amplify wind speed variability because of the "near" cubic dependence between power and wind speed. On the other hand, for wind speeds below cut-in wind speed or between nominal and cut-out wind speed, power fluctuations are smoothed considerably. However, if wind speed crosses cut-out speed or oscillates around it, fluctuations can be significantly increased as power varies from 100% (rated output power) to 0% (when the wind power plant is disconnected above cut out wind speeds). Figure 6 shows two different series of power: the reduction or increasing of variability depending on the wind speed range within the wind turbine power curve.

The second stage considers the reduction of the power fluctuations when aggregating several wind turbines within a wind power plant and even aggregating wind power plants in a large region. Aggregated power fluctuations are reduced by the diversity of the wind speeds within a large area. Analogously to the effect mentioned above, with regard to the rotor disk effect, spectral coherence can also be studied in wind power plants or even larger regions, analyzing which parts of the power fluctuations are not correlated or even which parts are delayed between wind turbines or power plants [19]–[20]. Studies on spectral coherence in wind power plants or regions can be found in the literature [21]–[23]. Examples of those effects are illustrated in Figure 7 and Figure 8, where power in a single turbine is compared with the aggregated power of a wind power plant and a set of wind power plants, respectively.

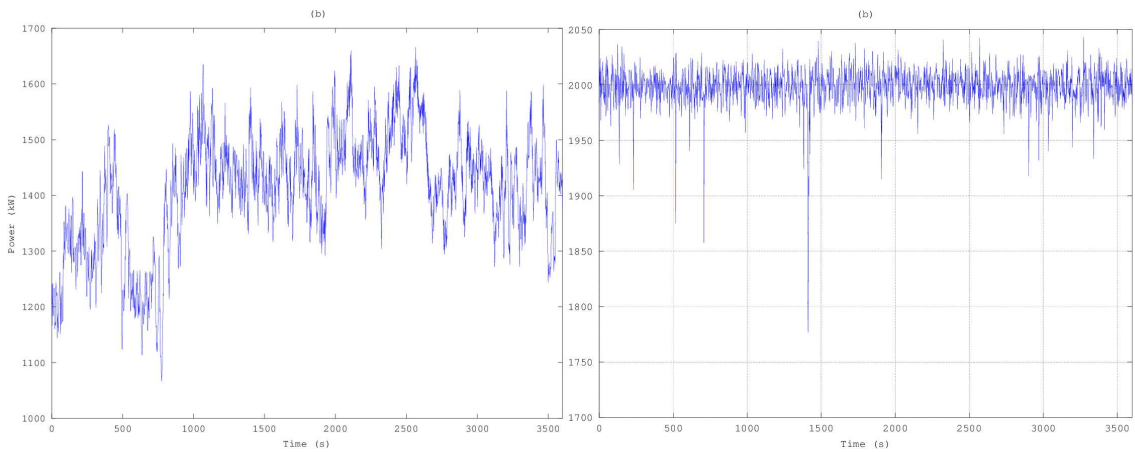

**Figure 6.** The figure on the left shows the power generated from a single wind turbine when wind is between cut-in and nominal speed; whereas the figure on the right shows wind speed between nominal and cut-out speed [23]

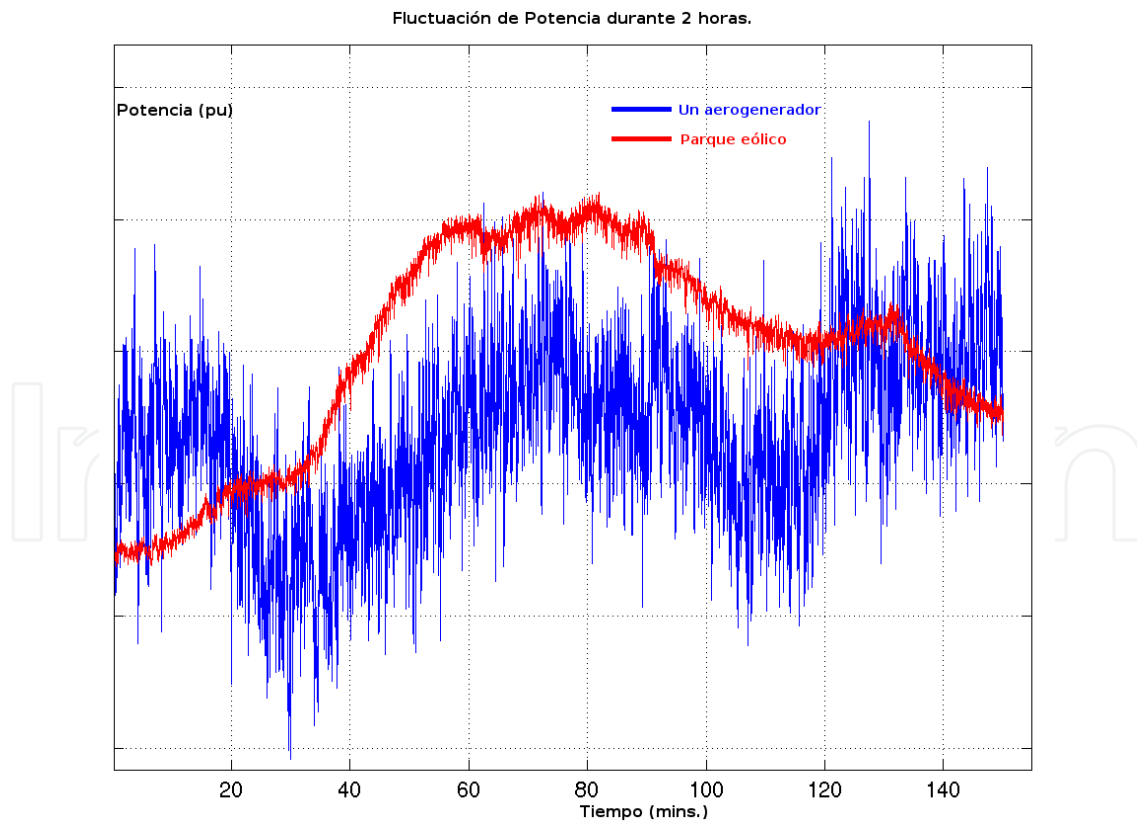

**Figure 7.** Comparison between a single wind turbine in an offshore wind power plant and the aggregated production of the whole wind power plant

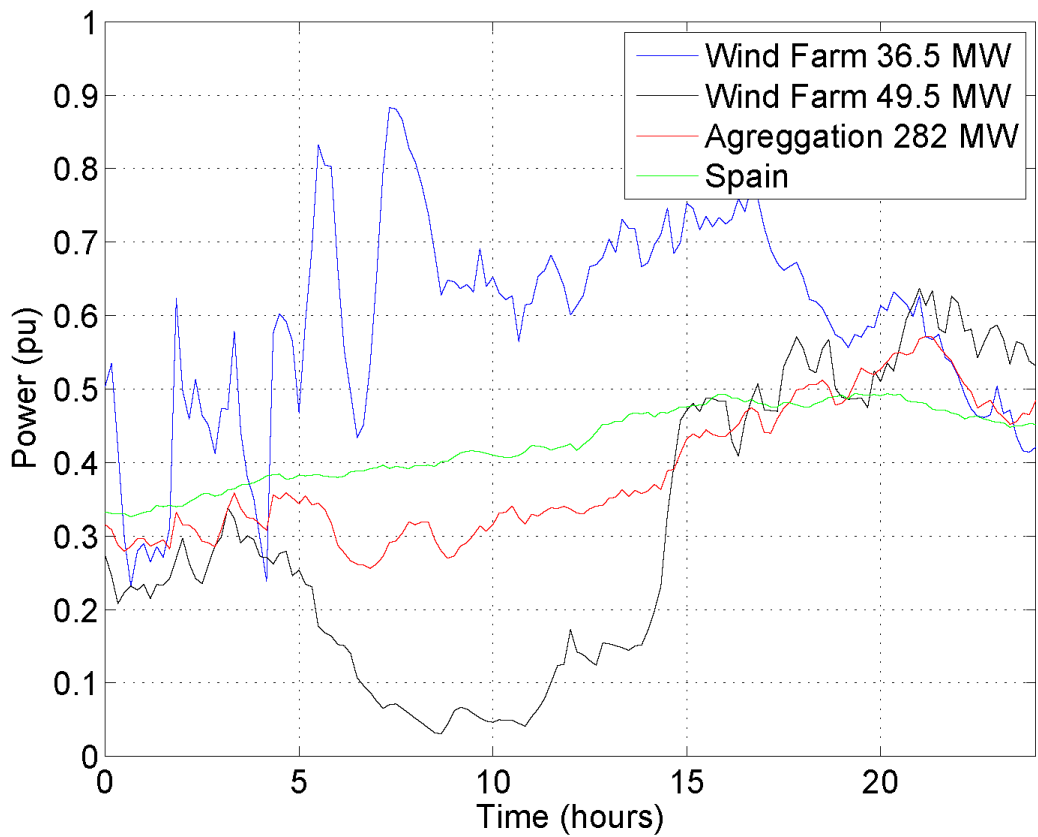

**Figure 8.** Comparison between the power produced by two wind power plants with the aggregation of nine wind power plants (including the previous ones) and the whole of Spanish wind power output

Finally, when aggregation includes different types of wind turbines, with different nominal wind speeds and different cut-out speeds, extreme aggregated fluctuations are also reduced significantly.In short, temporal diversities and spatial diversities reduces the peak-to-peak magnitudes of power fluctuations.

On the other hand, some regions wind power fluctuations are not related wind speed variability. Instead, the power fluctuations are caused by technical and operational challenges rather than by meteorological phenomena.Particularly, these include:

- Voltage sags. Voltage sags produce a sudden drop of wind power generation. This drop is usually recovered is quickly.

- Wind power curtailment. This is due to integration issues in the power system, such as limitations on the transmission or distribution networks, inability to ramp up or ramp down other generation sources, lack of enough reserves, etc.

Another classification is laid down. Attending to their ramping characteristics, wind power fluctuations events can be classified in:

- Wind power die-out. A wind power die-out refers to a persistent drop in wind power.

- Wind power rise. A wind power rise consists in a sustained rise in wind power that can create a persistent ramp up.

- Wind power lull. Wind die-outs are inevitably followed by wind rises. When both events happen in short succession, they form a wind power lull or a wind power dip.

- Wind power gust. A wind gust is opposite of a wind lull: it starts with a ramp up and ends with a ramp down.

## 4. Overview of spanish experience dealing with wind power variability: Examples of singular events

In this section, examples of the Spanish experience of singular events produced by either wind speed variability or operational issues are examined.

### 4.1. Voltage sags

Wind turbine manufacturers are required by transmission system operators (TSOs) to equip their turbines with fault ride-through (FRT) capability as the penetration of wind energy in the electrical systems grows [25]. Spain developed a procedure to measure and to evaluate the response of wind turbines and wind power plants subjected to voltage sags [26].The procedure for verification, validation, and certification of the requirements are described in the PO 12.3. This wind power plant commissioning and validation are based on the response of wind power plants in the event of voltage sags. The result of wind power plant commissioning leads to the certification of its conformity with the response requirements specified in the Spanish grid code [27]. Some aspects related to that grid code are explained in detail in [28]–[29].

On the other hand, because of the growing impact on power grid operations, the recent rapid expansion of wind generation has given rise to widespread interest in field testing and commissioning of wind power plants and wind turbines. Validation of computer dynamic models of wind turbines is not a trivial issue. Validation must ensure that wind turbine models represent with sufficient accuracy the performance of the real turbine, especially during severe transient disturbances [30]. In [32], different field tests for model validation and standards compliance are categorized according to the main input or stimulus in the test—control stimulus and external physical stimulus. Among these tests, the FRT capability of wind turbines can be performed using factory tests, at the individual wind turbine generator terminals, and using short-circuit field measurement data based on operational wind turbines and wind power plants.

Short-circuit field measurement data on operational wind turbines and wind power plants [33]—called opportunistic wind power plant testing in [31]—is performed with measurement equipment installed at the wind power plant site. The equipment records naturally occurring power system disturbances that are then used to validate wind turbine models. Power system modeling during the disturbances must be taken into account in the validation of wind turbine models. Therefore, monitoring wind power plants and wind turbines can be of interest for turbine manufacturers, wind power plant operators, and TSOs.Both the pre fault and the post fault data and power system network must be represented properly.

An extreme event recorded in Spain related to voltage sags occurred on March 19 and 20, 2007.Within twelve hours, four different disconnections of large amounts of wind power because of voltage sags were recorded. Those voltage sags were located in areas with high penetration wind power and during high wind speed periods. Figure 9 shows the recorded Spanish wind power output during these events.

The amount of wind power generation disconnected during these voltage sags were 553 MW, 454 MW, 989 MW, and 966 MW, respectively.

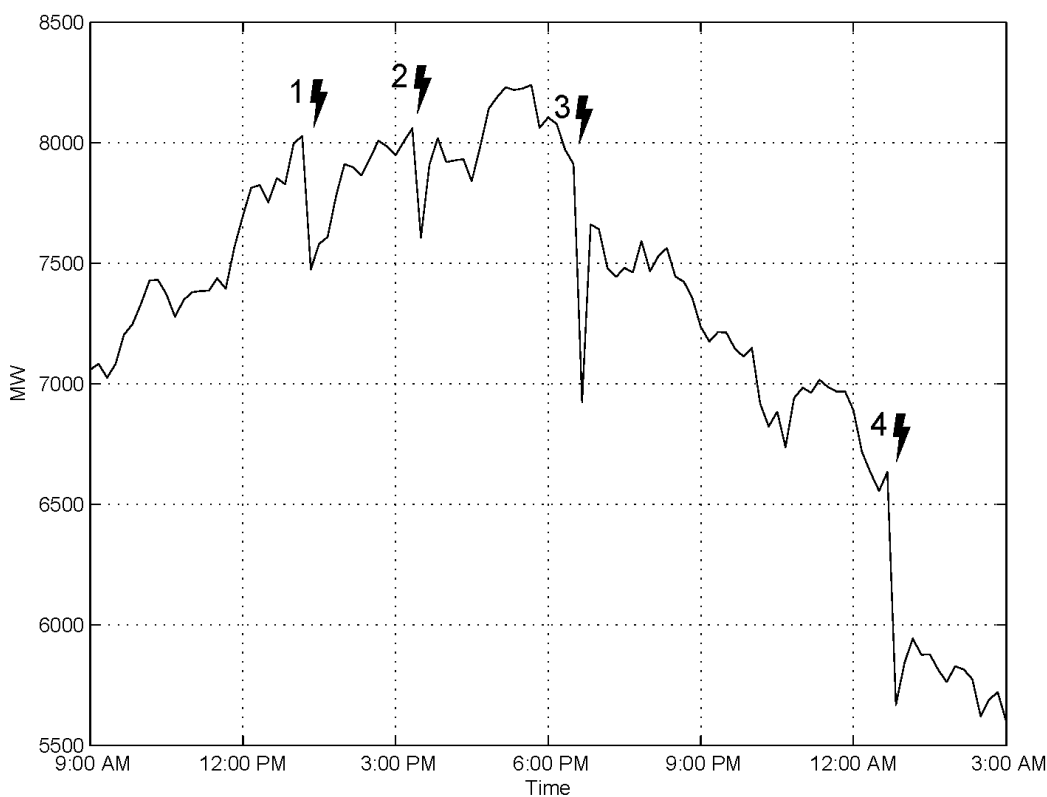

**Figure 9.** Spanish wind power during the voltage sags on March 19 and 20, 2007

In addition, nine Spanish wind power plants located in different areas were also analyzed during these events. Nominal power of these wind power plants varied from 6.8 MW at Wind Power plant 9 to 49.5 MW at Wind Power plant 4. In Figure 10, the wind power output from these nine wind power plants are presented. Highlights include:

- Wind Power plants 1, 2, and 3 are located in the same area. They were at high fluctuating partial load. These power plants were not affected by voltage sags because they were far away from the faults.

- Voltage Sags 1 and 2 affected only Wind Power plant 4.

- Voltage Sag 3 affected Wind Power plants 5, 6, and 7. These wind power plants are nearby. In these three cases, the responses to the sag were similar.

- Only Wind Power plant 9 was affected by Voltage Sag 4.

- All voltage sags during this period were located in areas with high wind power penetration.

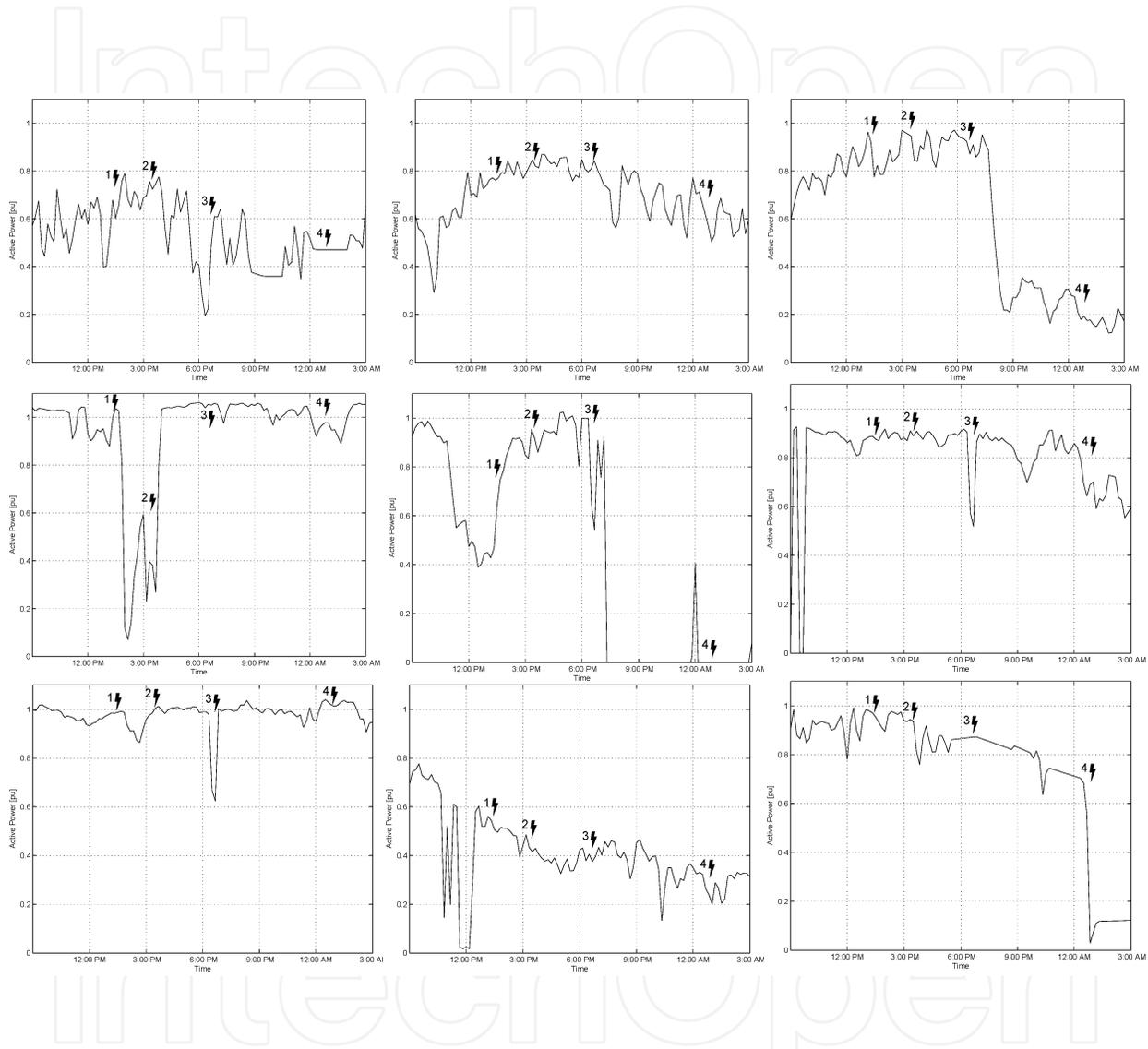

**Figure 10.** Wind power production of nine wind power plants during voltage sags on March 19 and 20, 2007

The operation of power systems under the effect of voltage sags in wind power has led TSOs to require FRT capability in wind power plants. By the end of 2010, 704 Spanish wind power plants had been certified against FRT capability (19.2 GW and around 95% of the installed capacity). A total of 1 GW wind turbines are excluded because of their missed manufacturers, small size, or because they are prototypes turbines. Figure 11 shows the number of power losses greater than 100 MW from 2005 and the percentage of wind power without FRT. As a result of the FRT implementations, the problem of significant wind generation tripping has been solved; therefore, wind plant curtailments have not been required since 2008.

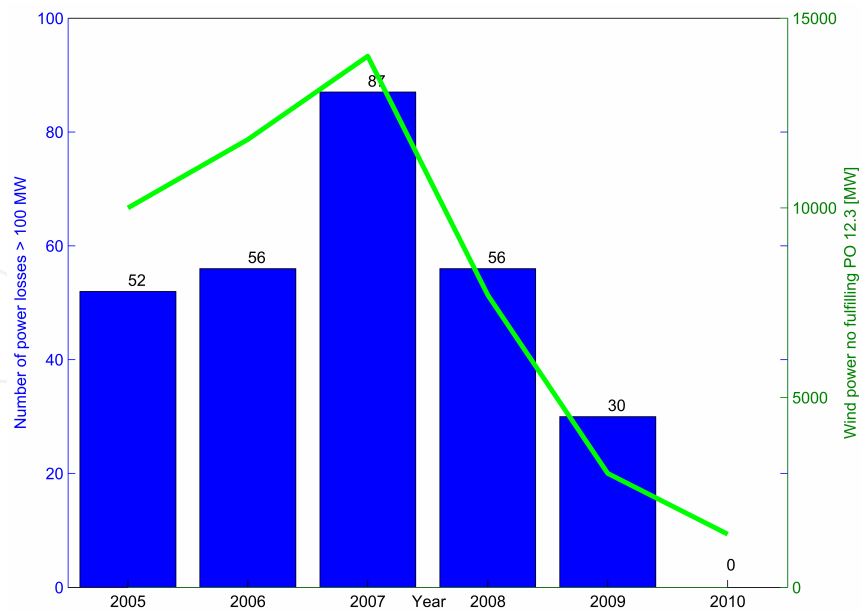

**Figure 11.** Evolution of wind power with FRT and number of power losses greater or equal to 100 MW by voltage sags in Spain [34]

The implementation of the supervisory control and data acquisition of wind generation in real time have decreased the number and the size of power curtailments, improved the quality and the security of the electricity supply, and maximized renewable energy integration. To further enhance wind energy integration, the Spanish TSO (Red Eléctrica de España, or REE) submitted a proposal of a new grid code (P.O 12.2) to the Ministry, with additional technical requirements for FRT, among others. The main purpose this proposal is to anticipate the expected problems in the Spanish power system between 2016 and 2020, by taking into account the incoming plants and new power plants to be deployed during these years to come. It is expected that P.O. 12.2 can be approved and applied in 2013.

### 4.2. Klaus Storm (January 23, 24, and 25, 2010)

Meteorological phenomena (e.g., storms or cyclones) are capable of causing large variations in wind power production and very high wind speeds. A storm within this category can affect a large number of wind turbines that have approximately the same cut-out wind speeds. When the cut-out speed is reached, the power generated goes from rated power to zero immediately. If this phenomenon spreads over several wind power plants in a particular area, it can cause a major threat to the power system stability and may lead to a cascading blackout.

The storm Klaus was named after an extra-tropical mid-latitude cyclone that struck between January 23 to January 25, 2009, affecting northern Spain and southern France. Wind speeds of higher than 150 km/h were recorded in the Spanish and French coastlines. The result was the disconnection of many wind power plants in northern areas of Spain, leading to a reduction of about 7,000 MW of wind power in a few hours (refer to Figure 12). Figure 13 shows the impact of the storm on a Spanish wind power plant. The wind power plant consists of 30 NEG Micon 82 Wind Turbines with a nominal power of 49.5 MW.

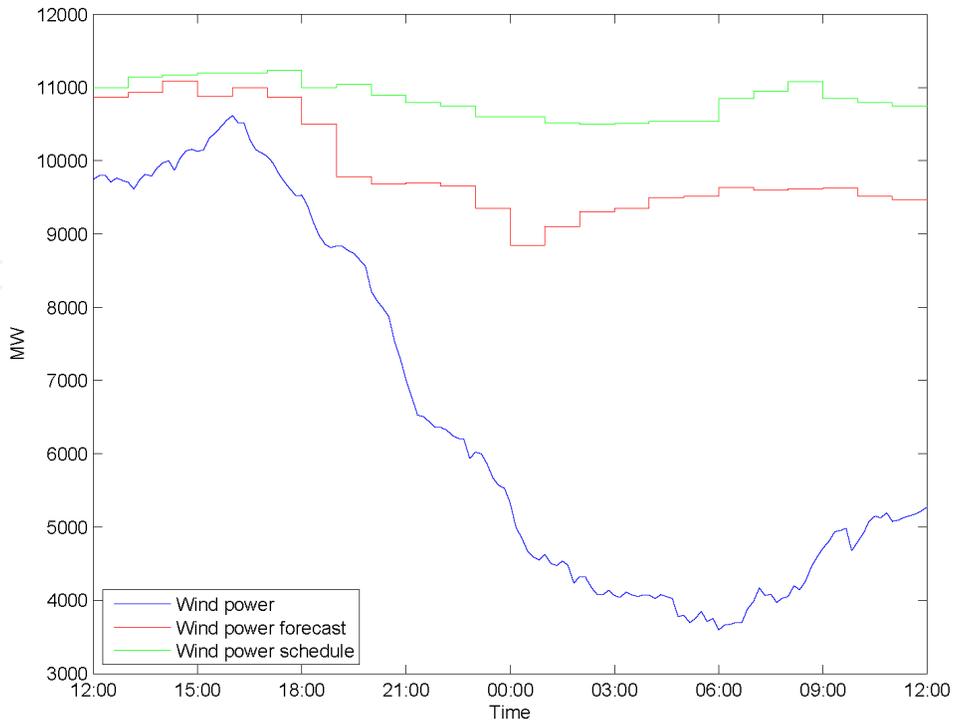

**Figure 12.** Wind power, forecasting, and schedule during the Klaus storm

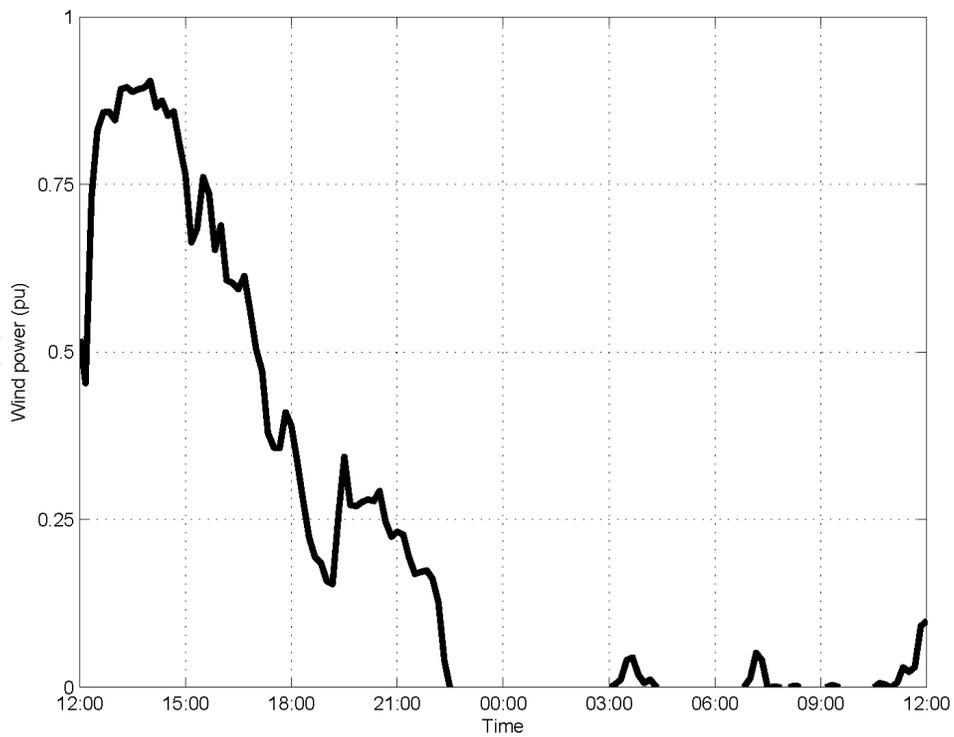

**Figure 13.** Wind power in a 49.5-MW wind power plant during the Klaus storm

During this emergency situation, the Spanish TSO REE has deployed several thermal power plants to increase the reserve power generation. Despite of the large difference between the forecast and actual wind power production, the power system continued to operate within the normal operating range.

This example shows the difficulties for forecasting wind power during these types of events.Differences between forecast and real wind power generation reached almost 6,000 MW. Furthermore, wind power decrement during the storm happened in the night, thus, it followed the ramping down of the daily load, so the increased reserves generation (6 GW) were sufficient to maintain the system balance.

### 4.3. Wind power curtailments

Wind energy curtailments because of integration issues in the power system have appeared in the Spanish power system. Until 2009, major curtailments were due to limitations on distribution networks, but since the end of 2009, cuts have been applied in real time to scheduled energy. However, the nature of renewable energy along with the economic and environmental issues, have provoked an interest in adding energy storage (such as pumped hydro storage PHS, a well-known technologies) into the power system mix.Spain accounts for around 5,000 MW (2.75 GW of pure PHS), with 77 GWh capacity. This technology is usually deployed because of the limited transmission capacity for exporting or importing power to neighboring countries.

As an example of wind power curtailment, Table 1 indexes orders delivered by the Spanish TSO on February 28, 2010. The initial and end times for every curtailment period are presented in columns 1 and 2. Column 3 represents Spanish wind power at the beginning of the period. In column 4, the Spanish TSO set point for this period is listed. In column 5, the real increase or reduction experimented by Spanish wind power in this period is shown. Finally, the ratio between the real increase/decrease and the increase/decrease obtained if wind power would match the set point is presented in column 6. In decrease periods, this ratio is equal to or higher than 1; in increase periods, it is equal to or less than 1, 1 being the optimum value.

| Initial Time | End Time | Wind Power (MW) | Wind Power Set Point (MW) | Real Increase/Reduction (MW) | Ratio |
|---|---|---|---|---|---|
| 1:08 | 2:07 | 7796 | 7331 | -796 | 1.71 |
| 2:07 | 3:48 | 6470 | 6099 | -718 | 1.93 |
| 3:48 | 6:08 | 5175 | 4904 | -720 | 2.66 |
| 6:08 | 8:45 | 4036 | 5904 | 217 | 0.11 |
| 8:45 | 9:10 | 3772 | 6905 | 420 | 0.14 |
| 9:10 | 9:43 | 3807 | 7905 | 276 | 0.07 |
| 9:43 | – | 4209 | Installed capacity | – | – |

**Table 1.** Curtailment schedule on February 28, 2010

### 4.4. Over-response to wind power curtailments

On January 1, 2010, the REE gave instructions for several wind power curtailments considering "Non-Integrable Wind Power Excess" as defined in Operational Procedure 3.7 [35]. During these curtailments, an over-response in the wind power plant power generation was obtained and the reduced power ratio was greater than four times the order required. This kind of event may threaten the power system operation, and from an economical point of view, because reserves generators are used for balancing, increasing costs are produced.

Figure 14 shows the sequence of curtailment instructions provided by the CECRE, the control center of renewable energies, together with the wind power generation in the power system. There were four orders with over-response during these hours, with effective wind power reduction from 2.42 to 4.02 times the commanded reduction.

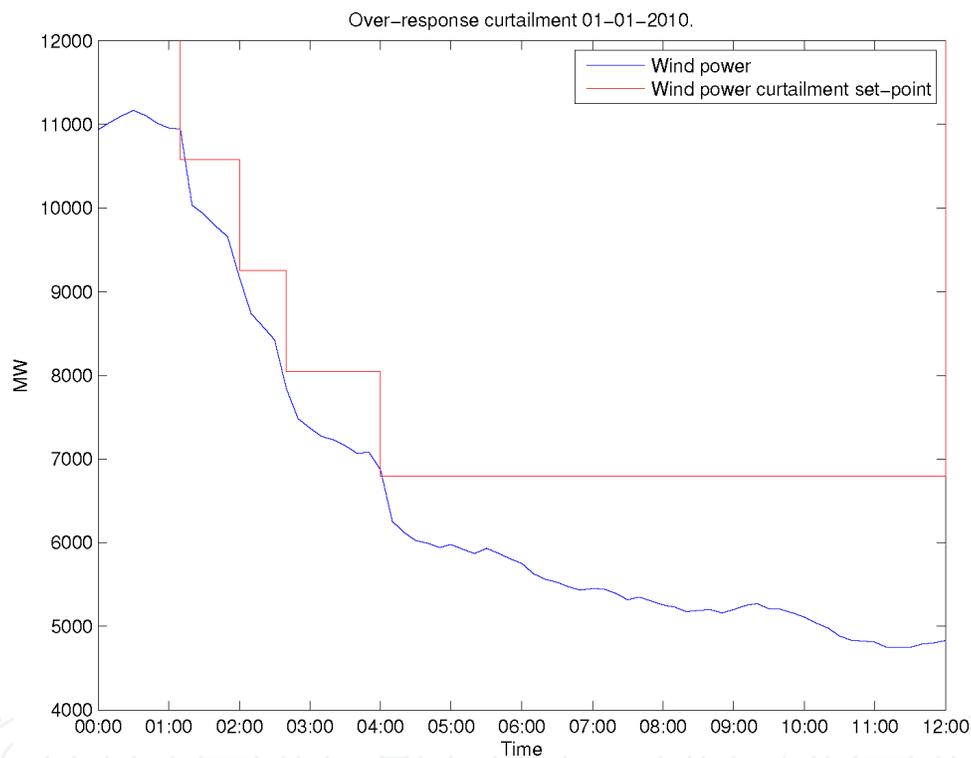

**Figure 14.** Over-response to curtailments in the entire Spanish wind power generation

The main causes of this over-response were:

- Curtailment is usually performed during low load and high wind penetration periods.

- During these periods, wind power plants often operate in high wind speed conditions and wind power fluctuations are dominant. Many wind power plants were stopped or operates in low output production due to cut-out protection disconnect the wind turbines operating above cut-out wind speeds.

- Curtailment is usually applied by disconnecting the entire wind power plant instead of turning off specific wind turbines (partial disconnections) within the wind power plant.

In Figure 14, an example of over-response to this curtailment is presented for the 49.5-MW wind power plant discussed previously. The TSO set point was ordered during early morning (03:00 to 07:00). When wind speed was above the cut-out wind speed (20 m/s), wind power decreased below the set point, reaching half generation and almost no generation. This additional drop must be replaced by the generation reserves.

The sequence of range of production was as follows:

- From 00:00 to 03:00, no curtailment was ordered. Most wind turbines were near or at 1 pu during this period. At 02:10, a slight wind power lull occurred as wind speed fell.

- From 03:00 to 04:40, a 0.6 pu TSO set point was applied. Then wind speed passed 20 m/s and most of the wind turbines were disconnected by cut-out speed protection. Wind power plant production fell to 0.1 pu, much less than 0.6 pu. Then wind speed went down, and wind power plant production almost reached the TSO set point. Some wind turbines were maintained at maximum available power; whereas others were disconnected, resulting in the TSO set point for the whole wind power plant. Some wind turbines were at maximum available power, and the rest remained disconnected. This kind of regulation involves repeated connections and disconnections during curtailment.

- From 04:40 to 07:40, the TSO set point changed from 0.6 pu to 0.5 pu. More wind turbines were disconnected to achieve this change.

- Finally, at 07:40, the TSO released the set point and the wind power plant recovered normal control.

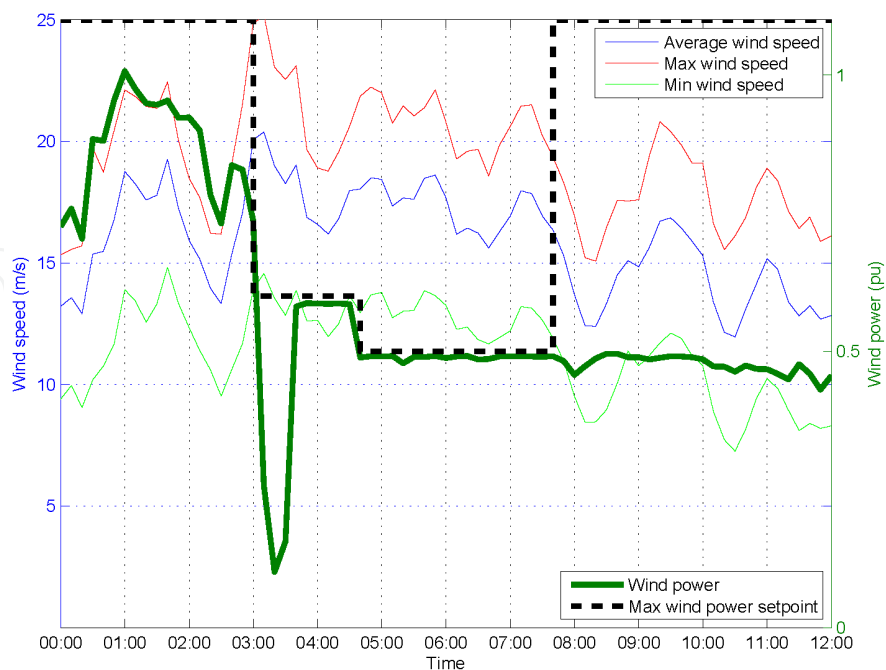

**Figure 15.** Example of an over-response in a 49.5-MW wind power plant on January 1, 2010

Possible solutions to avoid over-response to wind power curtailments, with the actual capacity of energy storage and transmission to other countries, are:

- The TSO should dispose of real-time wind power generation as well as a wind power forecast during the curtailment period. Maintenance schedules and cut-out shutdowns must be taken into account.

- Performing curtailment orders at the control center level instead of the wind power plant level must be studied.

- Control centers are connected to the CECRE and could improve curtailment management.

- Information about the reasons of curtailment, the application method, and wind power plant response could help toward overall optimization.

## 5. Conclusions

There are different types of events in which wind power fluctuates significantly. In some cases, fluctuations are produced by variations of wind speed, especially during meteorological events such as storms. Other power fluctuations are not directly linked with wind behavior and have a technical cause related to power system operation issues.

In this chapter, examples of the different events affecting wind power fluctuations were shown. The behaviors and the responses of the Spanish power system and wind power plants experiencing such events were analyzed. Examples presented in this chapter show that some of the wind power integration issues are related to low-voltage ride-through. They are solved through strict grid code enforcement. Other solutions to manage the reserve power generation and the wind power fluctuations are also very important in order to achieve high levels of wind power penetration. In the Spanish case, this could require increasing the availability of dispatchable and fast-start power plants, as well as increasing wind power plant participation on supporting the power system by providing voltage control, inertial emulation, frequency control, oscillation damping, or updated voltage ride-through capabilities.

## Acknowledgements

This work was supported in part by the "Ministerio de Ciencia y Innovación" (ENE2009-13106) and in part by the "Junta de Comunidades de Castilla-La Mancha" (PEII10-0171-1803), both projects co-financed with FEDER funds.

This work was also supported by the U.S. Department of Energy under Contract No. DE-AC36-08-GO28308 with the National Renewable Energy Laboratory.

## Author details

Sergio Martin-Martínez[1], Antonio Vigueras-Rodríguez[2], Emilio Gómez-Lázaro[1], Angel Molina-García[3], Eduard Muljadi[4] and Michael Milligan[4]

1 Renewable Energy Research Institute, Universidad de Castilla-La Mancha, Albacete, Spain

2 Renewable Energy Research Institute, Albacete Science and Technology Park and Universidad de Castilla-La Mancha, Albacete, Spain

3 Department of Electrical Engineering, Universidad Politécnica de Cartagena, Cartagena, Spain

4 National Renewable Energy Laboratory, Golden, Colorado, USA; Michael Milligan, National Renewable Energy Laboratory, Golden, Colorado, USA